\documentclass[twocolumn]{aastex62}
\usepackage{graphicx}
\usepackage{soul}

\graphicspath{{./}{figures/}}

\received{April 28, 2023}
\revised{X X, 2023}
\accepted{X X, 2023}

\submitjournal{ApJ}

\newcommand{\tablebib}{\tablecomments}

\newcommand{\msun}{\mbox{$M_\odot$}}
\newcommand{\mstar}{\mbox{$M_{\rm stellar}$}}
\newcommand{\mdust}{\mbox{$M_{\rm dust}$}}

\newcommand{\kms}{\mbox{km\,s$^{-1}$}}

\shorttitle{dusty ellipticals}
\shortauthors{Le\'sniewska et al.}

\begin{document}

\title{The Fate of the Interstellar Medium in Early-type Galaxies. II. Observational Evidence for Morphological Quenching
\footnote{Herschel is an ESA space observatory with science instruments provided by European-led Principal Investigator consortia and with important participation from NASA.}}

\correspondingauthor{Aleksandra Leśniewska}
\email{aleksandra.lesniewska@amu.edu.pl}

\author{Aleksandra Le\'sniewska}
\affiliation{Astronomical Observatory Institute, Faculty of Physics, Adam Mickiewicz University, ul. S{\l}oneczna 36, Pozna{\'n}, Poland}
\affiliation{DARK, Niels Bohr Institute, University of Copenhagen, Jagtvej 128, DK-2200 Copenhagen N, Denmark}

\author{M.~J.~Micha{\l}owski}
\affiliation{Astronomical Observatory Institute, Faculty of Physics, Adam Mickiewicz University, ul. S{\l}oneczna 36, Pozna{\'n}, Poland}

\author{C.~Gall}
\affiliation{DARK, Niels Bohr Institute, University of Copenhagen, Jagtvej 128, DK-2200 Copenhagen N, Denmark}

\author{J.~Hjorth}
\affiliation{DARK, Niels Bohr Institute, University of Copenhagen, Jagtvej 128, DK-2200 Copenhagen N, Denmark}

\author{J.~Nadolny}
\affiliation{Astronomical Observatory Institute, Faculty of Physics, Adam Mickiewicz University, ul. S{\l}oneczna 36, Pozna{\'n}, Poland}

\author{O.~Ryzhov}
\affiliation{Astronomical Observatory Institute, Faculty of Physics, Adam Mickiewicz University, ul. S{\l}oneczna 36, Pozna{\'n}, Poland}

\author{M.~Solar}
\affiliation{Astronomical Observatory Institute, Faculty of Physics, Adam Mickiewicz University, ul. S{\l}oneczna 36, Pozna{\'n}, Poland}

\begin{abstract}

The mechanism by which galaxies stop forming stars and get rid of their interstellar medium (ISM) remains elusive. Here, we study a sample of more than two thousand elliptical galaxies in which dust emission has been detected. This is the largest sample of such galaxies ever analysed. We infer the timescale for removal of dust in these galaxies and investigate its dependency on physical and environmental properties. We obtain a dust removal timescale in elliptical galaxies of $\tau$ = 2.26 $\pm$ 0.18 Gyr, corresponding to a half-life time of 1.57 $\pm$ 0.12 Gyr. This timescale does not depend on environment, stellar mass or redshift. We observe a departure of dusty elliptical galaxies from the star formation rate vs.~dust mass relation. This is caused by the star-formation rates declining faster than the dust masses and indicates that there exists an internal mechanism, which affects star formation, but leaves the ISM intact. Morphological quenching together with ionisation or outflows caused by older stellar populations (supernova type Ia or planetary nebulae) are consistent with these observations.

\end{abstract}


\section{Introduction}

Dust influences the evolution of galaxies by acting as catalyst of molecule formation and providing shielding from interstellar radiation. Its emission can also be used as a diagnostic for interstellar medium (ISM) properties \citep{Scoville2016}. There are several processes that can contribute to dust removal from galaxies. Dust can be incorporated in newly formed stars (astration; \citealt{Gall2018}), or destroyed by active galactic nucleus (AGN) feedback \citep{Fabian2012}.  Supernovae (SNe) may destroy newly-formed and pre-existing dust by forward and reverse shock waves \citep{Temim2015,Bianchi2007,Cherchneff2010,Gall2011,Lakicevic2015}.
Dust can be also destroyed by planetary nebulae. This is due to heating of gas by shocks from colliding  planetary nebulae \citep{Conroy2015}. Galactic outflows contribute to dust removal and can be very effective due to radiation pressure-driven dusty flows
\citep{Bianchi2005}. 
Hot gas ($\sim$ 10$^{5}$ K) present in some regions of ISM can also cause erosion of dust particles. The smallest grains are the most vulnerable to this mechanism \citep{Bocchio2012}. 

Over the past decades, many theoretical works have been developed to model the formation, evolution and destruction of dust in galaxies.
Among the first research dealing with dust evolution is \citet{Dwek1980}, who emphasized the importance of SNe. \cite{Barlow1978} studied sputtering of dust grains in \ion{H}{2} regions, inter-cloud medium, cloud-cloud collisions shock waves, and SN remnants, concluding that the latter dominates this process.  
\cite{Gall2011} developed a numerical model of galactic chemical evolution and studied the effect of galaxy properties on the evolution of dust. Dust destruction was described in the model as being caused by SN shocks. The tested properties of dust evolution depend very strongly on the initial mass function. 
\cite{Slavin2015} focused on dust destruction by SNe, which resulted in a dust removal timescale of 2--3\,Gyr.

Recent studies of high-redshift ($z\sim 1.6$--3.3) lensed quiescent galaxies have shown that their dust-to-stellar mass ratios are of order 10$^{-4}$ \citep{Whitaker2021}. Similarly, \cite{Blanquez2023} showed that high-redshift galaxies are characterized by an order of magnitude higher gas fractions than what is detected in the local universe.

In order to separate the processes of dust formation and removal, it is an advantage to study galaxies with little dust formation, but with detectable ISM. Therefore,  dusty early-type galaxies (ETG; ellipticals and lenticulars) form a suitable sample for such endeavour. 
The dust emission of only several dozen of such galaxies has been analysed \citep{Smith2012,Rowlands2012,Agius2013,Agius2015,diseregoalighieri2013,Hjorth2014,Dariush2016,Michalowski2019,Magdis2021}. 
\cite{Hjorth2014} showed that dusty early-type galaxies do not follow the relation between the star formation rates (SFRs) and dust masses \citep{Cunha2010} and discussed formation or quenching scenarios. \cite[][submitted 2023]{Michalowski2019} revealed an exponential decline of the dust-to-stellar and gas-to-stellar mass ratios with galaxy age and measured the timescale of this process to be 2.5 $\pm$ 0.4\,Gyr. To date, this is the only measurement of the dust removal timescale in dusty early-type galaxies and is based on a sample of 61 galaxies. 

Dusty elliptical galaxies are quite rare. Hence, far-infrared/submillimeter surveys need to cover a large area to detect a high number of  galaxies to build a significant sample. The ESA \textit{Herschel Space Observatory} (henceforth \textit{Herschel}, \citealt{Pilbratt2010}) has provided deep infrared observations of hundreds of square degrees of the sky. Its large field of view, 4$'$ $\times$ 8$'$, and sensitivity has led to the detection of dust in millions of galaxies.

One of the major cosmological and galaxy evolution observation projects, Galaxy And Mass Assembly (GAMA; \citealt{Driver2011,Driver2016,Baldry2018,Smith2011}\footnote{\url{http://www.gama-survey.org}}), brings together the latest generation of instruments and surveys, such as the Anglo-Australian Telescope (AAT), Sloan Digital Sky Survey (SDSS), and {\it Herschel}. 
These datasets were combined in a database of several hundred thousand galaxies, with a magnitude limit in the $r$ band of $19.8$ mag.
Such an extensive catalog not only allows the examination of the relationships between individual quantities, but also gives the possibility of additional sampling into bins of various parameters. 

In this paper we study a large sample of more than two thousand elliptical galaxies in which dust was detected. The sample size allows us to investigate dust evolution as a function of various galaxy properties. We focus on relationships between their physical and environmental parameters. The objective of this paper is to distinguish the mechanisms contributing to the removal of dust in elliptical galaxies and investigate its dependency on physical and environmental properties.

We use a cosmological model with $H_0$ = 70 km s$^{-1}$ Mpc$^{-1}$, $\Omega_{\Lambda}$~=~0.7, and $\Omega_m$~=~0.3. We assume the \citet{Chabrier2003} initial mass function.

\section{Data and Sample}

\subsection{GAMA Catalog}

{\it Herschel} covered an area of 161.6 deg$^2$ of the GAMA fields and provided information on dust emission at 250, 350, and 500 $\mu$m.  
The GAMA catalog for these fields contains properties of 120,114 galaxies based on modeling of spectral energy distributions with the Multi-wavelength Analysis of Galaxy Physical Properties (MAGPHYS; \citealt{daCunha2008}). This includes dust masses, stellar masses, star formation rates, and luminosity-weighted stellar ages. The values of these parameters were obtained by the GAMA project and are presented in their MagPhys catalogue\footnote{\url{http://www.gama-survey.org/dr3/data/cat/MagPhys/}}.
We also obtained a wide range of parameters related to photometry, a single-S\'ersic fit to SDSS 2D surface brightness distribution \citep{Kelvin2012} and local environment of galaxies such as surface galaxy density ($\Sigma$) calculated based on the distance to the fifth nearest neighbour within a velocity difference of $\pm$1000~{\kms} \citep{Brough2013}.

\subsection{Sample}

We used the r-band S\'ersic index \citep{Sersic1963}, $n$, to select elliptical galaxies by requiring that $n>4$. This resulted in 22,571 galaxies.

From this set of galaxies we selected dusty ellipticals with a minimum signal-to-noise ratio at the {\it Herschel} SPIRE \citep{spire} $250\,\micron$ filter of 3. This step resulted in 2,956 galaxies, so 13\% of elliptical galaxies are detected by {\it Herschel}. This is higher than the detection rate of 5.5\% obtained by \citet{Rowlands2012} for similar galaxies, who required a higher significance of $5\sigma$ at $250\,\micron$.

\cite{Rowlands2012} visually classified galaxies to the early-type category at redshifts $0.01 < z < 0.32$. We selected galaxies in the same redshift range. At higher redshifts the morphological classification is uncertain \citep{deAlbernazFerreira2018} and the sample could contain compact high star-forming (not elliptical) galaxies. Our final selection, including the redshift cut, resulted in 2050 galaxies. Our selection roughly corresponds to a flux-limited sample above 20.7 mJy at the SPIRE $250\,\micron$, although adopting that limit would result in 17$\%$ of galaxies having a signal-to-noise ratio less than 3. Selection of galaxies based on SPIRE $250\,\micron$ flux $>$ 20.7 mJy does not affect our results.

The uncertainties of the physical properties are the following, measured separately for MS and below-MS subsamples:
0.12--0.14 dex for stellar age, 0.1--0.3 dex for SFR, 0.15--0.22 dex for M$_{dust}$, 0.1 dex for M$_{stellar}$, where the higher values correspond to the galaxies below the main sequence.

\citet{Rowlands2012} estimated that 2\% of dusty early-type galaxies in their sample are likely chance projections of a dust-free galaxy and a background dusty galaxy. Our selection criteria are similar: we used the updated GAMA archive (DR3), S\'ersic index $>$ 4 instead of visual classification, and the same redshift range, so we expect a similar fraction, which does not affect our analysis. The main difference is the area over which the galaxies were selected, resulting in a much larger sample of 2050 objects as compared to the 44 galaxies studied in \cite{Rowlands2012}.

\section{Results}

\subsection{Main Sequence}

We divided the selected galaxies into two groups: galaxies within and below the main sequence (MS) of star forming galaxies. Fig.~\ref{Fig1} (top) presents a comparison of our galaxies with a redshift-dependent MS as measured by \citet[eq. 28]{Speagle2014}. We adopted a measured MS width of 0.2 dex \citep{Speagle2014}, independent of redshift. Any galaxy below the MS by more than 0.2\,dex is assigned to the `below-MS' group in this paper. Our sample covers the redshift range uniformly with a sensitivity $<$ 100 times below the MS at all redshifts. This resulted in 722 MS dusty elliptical galaxies and 1\,328 below-MS galaxies.

We tested the validity of the Speagle MS for our data using late-type galaxies from GAMA, which have been selected based on S\'ersic index $<$ 2.5, 0.1 $<$ z $<$ 0.15, and S/N $>$ 3 at S250. We find an agreement between the Speagle MS and the MS estimated using the selected LTGs, in particular in the stellar mass range covered by our ETG sample.

\subsection{Dust Removal Timescale}

Figure \ref{Fig1} presents dust-to-stellar mass ratio as a function of luminosity-weighted stellar age (middle panel). There is an evident decline in the mass ratio as galaxies evolve over time. Fitting an exponential function to this plane, as in \cite{Michalowski2019}, allows us to evaluate the timescale of the dust mass removal for different galaxy properties:
\begin{equation}\label{exponential}
    \frac{M_{dust}}{M_*} = A \cdot e^{-age/\tau},
\end{equation}
where $A$ is the normalisation constant and $\tau$ is the dust removal timescale. We obtained a dust removal timescale for all elliptical galaxies of $\tau=2.26 \pm 0.18$~Gyr with the corresponding half-life time of $1.57 \pm 0.12$~Gyr. The values of the dust removal timescale, half-life time and the normalisation constant are presented in Table~\ref{tab1}. To our knowledge, this is the first determination of the dust removal timescale for such a large sample and for different galaxy properties. 

We also fit the exponential function separately to galaxies on and below the MS. 
The elliptical galaxies below the MS (red line) follow the fit obtained by \cite[][lime green line]{Michalowski2019}, whereas the elliptical galaxies on the MS (blue line) are characterized by a faster dust mass decline. The results of our fitting are given in Table \ref{tab1}.

One of the basic parameters which is useful for subdivision into smaller bins is stellar mass, because galaxies of different masses may evolve differently. The three top panels in Fig.~\ref{Fig2} show the dust-to-stellar mass ratio as a function of age for three stellar mass bins between $10<\log(M_{\rm stellar}/M_\odot)< 11.5$, with a 0.5 dex width. The fits for these stellar mass bins are consistent with each other within the error bars (Table~\ref{tab1}). Therefore, we conclude that the dust mass decline with time does not depend on stellar mass in the analysed range. 

The most massive group with $11.5 < \log(M_{\rm stellar}/M_\odot) < 12.2$, does not contain MS galaxies, and includes only galaxies with high ages and low dust-to-stellar mass ratios. It is not possible to fit an exponential function to the galaxies in this group because the dynamical range of both properties is too small. However, these galaxies are still consistent with the fitted dust removal function obtained for galaxies at lower masses. 

Other galaxies in the close proximity of elliptical galaxies can affect their ISM. Therefore, we studied the role of the galaxy environment. The GAMA catalog provides surface galaxy density, $\Sigma$, in the G15 field for galaxies at $z<0.18$ \citep{Brough2013}. There are 384 of our dusty ETGs satisfying these criteria and for 373 of them (97\%) $\Sigma$ has been measured. The dust decline as a function of age in bins of $\Sigma$ is presented in Fig.~\ref{Fig2} (middle row). It is evident that the decline in dust mass is independent of the galaxy environment. We reached the same conclusion when we analysed the effect of environment in narrower ranges of stellar mass.

Our sample spans a redshift range $0.01$--$0.32$, corresponding to 3.6~Gyr of the evolution of the Universe. Fig.~\ref{Fig2} (bottom) shows that the dust removal does not depend on redshift, as galaxies follow the same dust removal trend at each redshift bin. 

\subsection{Dust Masses vs Star Formation Rates}

\begin{figure*}
\includegraphics[width=\textwidth]{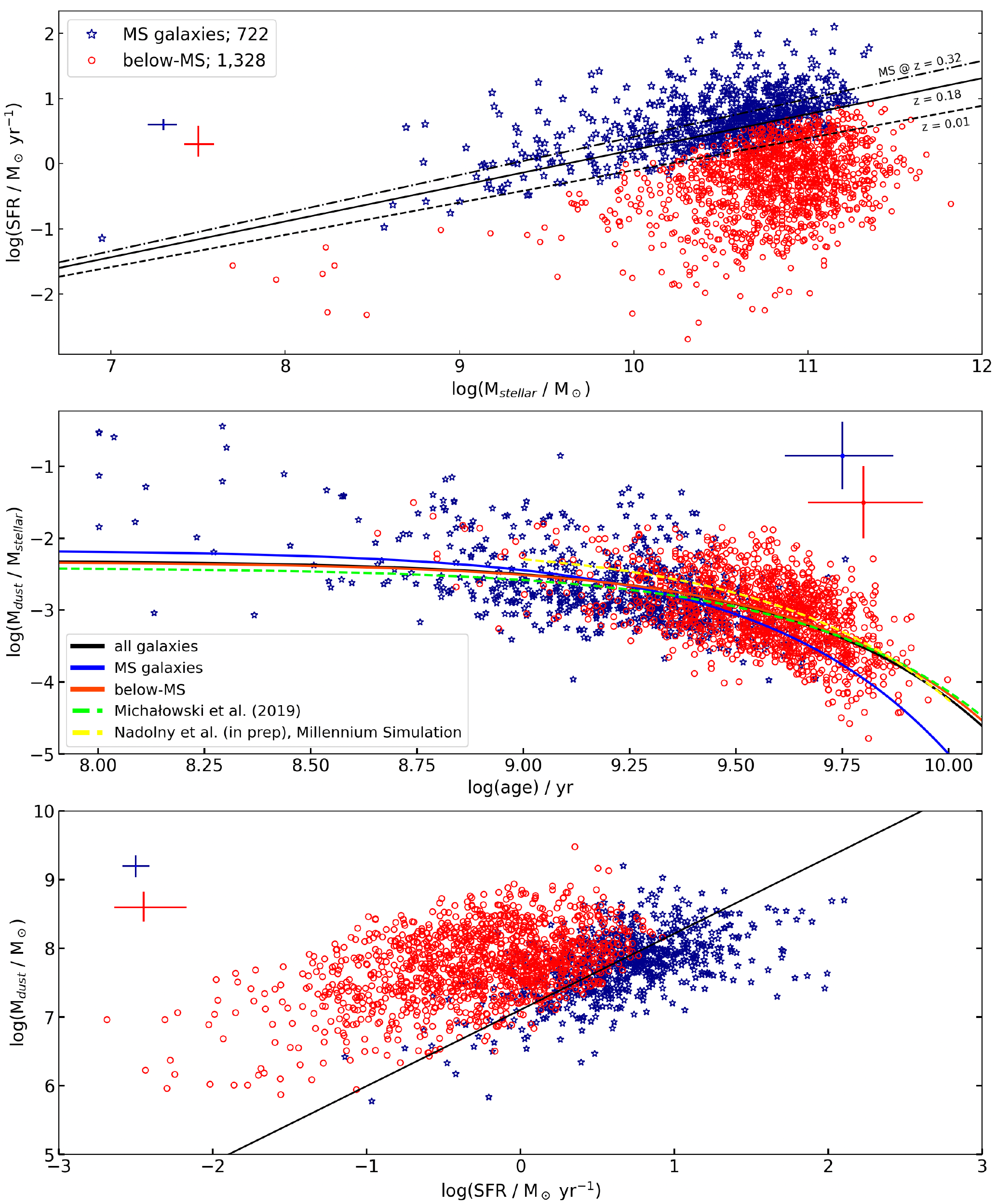}
\caption{(Top) SFR as a function of stellar mass. Color coding distinguishes MS galaxies (blue stars) and galaxies below the MS (red circles). The star formation main sequence at $z$ = 0.32, 0.18, and 0.01 (black lines) based on \cite{Speagle2014} are shown. The numbers of selected MS and below-MS galaxies are presented in the legend. (Middle) Dust-to-stellar mass ratio as a function of stellar age. The exponential fits are for galaxies within the MS (blue line), galaxies below the MS (red line), all galaxies (black line), that obtained by \citet{Michalowski2019} (lime green dashed line), and by \cite{Nadolny2023} with the Millenium Simulation (yellow dashed line) within the age range 9.0-- 10.1 Gyr. (Bottom) Dust mass as a function of SFR with the \cite{Cunha2010} relation (black line). Median errorbars for the MS and below-MS galaxies are shown as blue and red crosses, respectively.}
\label{Fig1}
\end{figure*}

\begin{figure*}
\includegraphics[width=\textwidth]{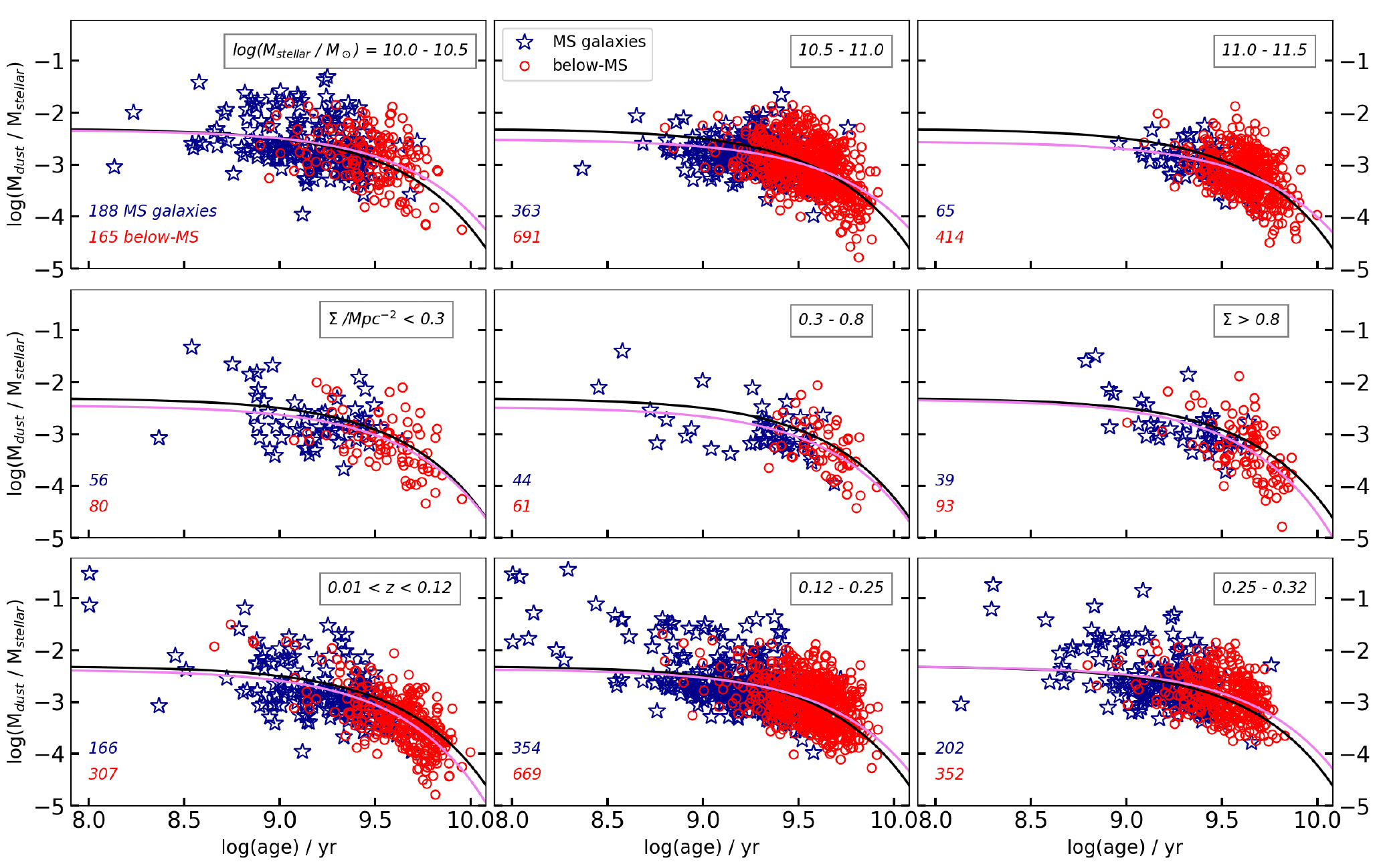}
\caption{Dust-to-stellar mass ratio as a function of stellar age and other galaxy properties. The MS galaxies are marked as blue stars and galaxies below the MS are marked as red open circles. The numbers of selected MS and below-MS galaxies in each panel are shown. The exponential fits are for all 2050 studied galaxies (black line) and for galaxies plotted on individual panel (violet line). The division into three stellar mass bins (top row), galaxy surface density based on the distance to the 5th nearest neighbour (middle row), and redshift (bottom row) are shown.}
\label{Fig2}
\end{figure*}

Figure \ref{Fig1} (bottom) presents the SFR-{\mdust} relation for our 2050 dusty elliptical galaxies. It is evident that our MS elliptical galaxies follow the \cite{Cunha2010} relation (black line). Hence for MS elliptical galaxies the decrease in SFR is accompanied by a similar decrease in the dust mass, so they stay on the relation. However, as first shown by \cite{Hjorth2014}, elliptical galaxies below the MS are found above the \cite{Cunha2010} relation with higher dust masses than what their SFRs imply.

\subsection{Central Surface Luminosity}
\label{sec:surf}

From the GAMA light profile catalog we used the values of the central surface brightness and converted them to central surface luminosities (luminosity per kpc$^2$). We find that the decrease of dust mass with the age of the elliptical galaxies does not depend on the central surface luminosity.

\begin{figure*}
\includegraphics[width=0.95\textwidth]{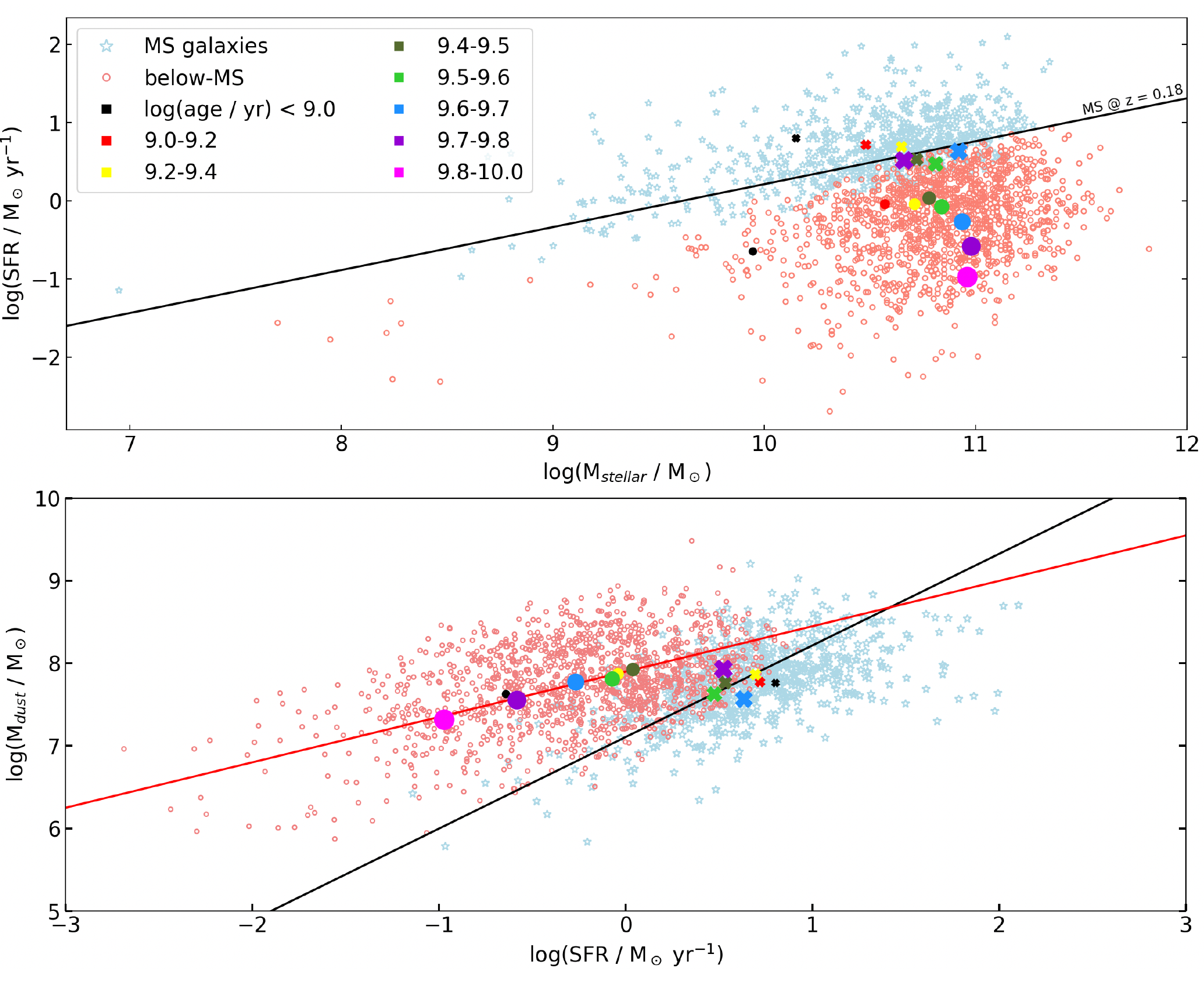}
\caption{SFR as a function of stellar mass (top) and dust mass as a function of SFR (bottom). Color coding distinguishes MS galaxies (blue stars) and galaxies below the MS (red circles). The star formation main sequence at $z$ = 0.18  based on \cite{Speagle2014} and the \cite{Cunha2010} relation are shown (black lines). The median values of SFR, stellar age, and dust mass for eight galaxy age ranges are marked as filled crosses for the MS galaxies, and as filled circles for galaxies below the MS. In addition to the color-coding, the size of the symbol increases with age. The red line shows a power-law fit to the median values of the galaxies below the MS in a form $\log(M_{dust}) = (0.55^{+0.10}_{-0.11}) \cdot \log(\mbox{SFR}) + (7.893^{+0.030}_{-0.031})$.}
\label{Fig3}
\end{figure*}

\subsection{Quenching}

To study the evolution of dusty elliptical galaxies, we divided our sample into eight bins of stellar age. Figure~\ref{Fig3} (top) presents SFR vs.~stellar mass with the addition of the median values in age bins, separately for the MS and below-MS elliptical galaxies. These medians are presented in Table~\ref{tab2} in the Appendix. The medians of SFRs and stellar masses of MS elliptical galaxies are (as expected) close to the MS. For elliptical galaxies below the MS, with increasing age the medians move away from the MS toward lower SFRs. The youngest below-MS elliptical galaxies are $\sim0.6$\,dex below the MS and the oldest have more than 10 times lower SFRs than the youngest.

Figure \ref{Fig3} (bottom) presents dust mass vs.~SFR with the medians in age bins for the MS and below-MS elliptical galaxies (Table~\ref{tab2}). The medians for MS ellipticals are located close to each other and to the \cite{Cunha2010} relation (black line), with no clear evolution. Elliptical galaxies below the MS have higher dust masses for their SFRs than what the \cite{Cunha2010} relation implies. We fitted a power-law function to the medians of the galaxies below the MS (red line), resulting in $\log(M_{dust}) = (0.55^{+0.10}_{-0.11}) \cdot \log(\mbox{SFR}) + (7.893^{+0.030}_{-0.031})$. This is shallower than the slope of the \cite{Cunha2010} relation of $1.1$. 
 
From Fig.~\ref{Fig3} (bottom) it is evident that the elliptical galaxies below the MS move away from the \cite{Cunha2010} relation as they are getting older. The youngest of the below-MS galaxies have SFRs around 1~$M_{\odot}$~yr$^{-1}$ and dust masses around 10$^{7.9}$~$M_{\odot}$ (0.8\,dex above the relation). With increasing age, their SFRs decrease faster than their dust masses. This results in the oldest galaxies having SFR around $0.1$~$M_{\odot}$~yr$^{-1}$ (a factor of 10 decrease) and dust mass of 10$^{7.3}$~$M_{\odot}$ (a factor of 4 decrease), placing them 1.3\,dex above the relation.

\subsection{Sample Evaluation}

To ensure that our selection is robust, we studied a subsample of galaxies with at least two detections among 5 {\it Herschel} bands (S/N $>$ 3). This resulted in 1430 galaxies. The exponential curve fitting gives the same results as the original sample within the error limits of these parameters. This shows that increasing the number of required band detections does not change our results and conclusions.

In order to check the correctness of the stellar ages calculated by the GAMA project, we analyzed average spectral energy distribution (SED) for eight stellar age bins defined in previous section. There is a clear correlation between the bin age and the relative normalised (in the near-infrared) flux. The oldest bin shows lower flux at the blue part of the SED, while the youngest bin shows the most prominent blue part of the SED that corresponds to the young stellar population of massive and hot OB stars. Normalisation in the near-infrared (equivalent to a stellar mass normalisation, as considered above), gives a clear luminosity decrease in the far-infrared with increasing age, equivalent to the dust-to-stellar decrease.

The GAMA project database also contains information about the D4000 break \citep{Cardiel1998, Balogh1999}. The strength of this break as a function of luminosity-weighted age for the below-MS galaxies from our sample shows that older galaxies have higher D4000, consistent with the determined age of the galaxies. The Spearman's rank correlation is 0.47 and the probability of the null hypothesis of no correlation is $10^{-70}$.

\section{Discussion}

Our key result is the confirmation of the exponential decrease of the dust mass with age using unprecedentedly large sample. We also found that SFRs of dusty ellipticals below the MS decline faster with age than their dust masses and the dust mass decline is independent of stellar mass, environment, redshift and central surface luminosity. As suggested by \cite{Hjorth2014} and \citet[][submitted 2023]{Michalowski2019}, morphological quenching is a potential mechanism for departing from the \cite{Cunha2010} relation. This is consistent with our findings. The process may be responsible for the gravitational stability that stops the collapse of gas clouds, resulting in a slower rate of star formation. At the same time, the process does not change the amount of gas, which means that the dust mass observed in these galaxies does not decrease proportionally with the SFR. Other processes must be responsible for the decline of the dust masses, e.g., the destruction of dust by feedback from older stellar populations (see Micha{\l}owski et al. submitted). This includes SNe Type Ia \citep{Li2020} or planetary nebulae \citep{Conroy2015}.

AGN feedback is also one of the potential mechanism of the ISM removal \citep{Fabian2012}. Recent studies suggest that quenching is connected with integrated AGN feedback over the lifetime of a galaxy, which is correlated with the supermassive black hole mass, not the instantaneous AGN luminosity \citep{Bluck2020,Bluck2020b,Bluck2022,Bluck2023,Piotrowska2022}. This mass is correlated with the bulge mass \citep{Magorrian1998,Haring2004}, which can be approximated by the galaxy central surface luminosity. We  did not detect any dependence on the dust decline on this parameter (Section~\ref{sec:surf}), which suggests that integrated AGN feedback is not a dominating mechanism of the dust removal. This is because if the integrated feedback was responsible for the dust removal in our galaxies then galaxies with higher central surface luminosities (more massive black holes and therefore stronger feedback) would exhibit a faster ISM decline. This finding is consistent with our study of the \citet*[][BPT]{bpt} diagram which shows that only up to 15\% of galaxies in our sample host AGNs, which means that they cannot have any significant effect on reducing the dust amount in these galaxies \citep{Ryzhov2023}.

We did not find any redshift dependency or environmental influence on dust removal, which is inconsistent with external mechanisms of dust removal.
The dust removal also does not depend on the stellar mass [in the explored range of $\log(\mstar/\msun)=10$--$11.5$], so the process linearly scales with mass (a bigger galaxy has proportionally more dust and proportionally 
more efficient dust removal).

We note that the lack of the below-MS elliptical galaxies at or even below the SFR-{\mdust} relation is not due to a detection limit at {\mdust}. It is $10^{5.2}\,\msun$ at $z = 0.05$ and $10^{6.7}\,\msun$ at $z = 0.3$ \citep{Michalowski2019}, so if such galaxies existed, they would be detected.

\section{Conclusions}

We analysed ISM and stellar properties of 2\,050 dusty elliptical galaxies which has never been done before on such a large sample.Our findings support the morphological quenching as a mechanism behind their SFR decline, as proposed by \cite{Hjorth2014}. This is because the  galaxies below the MS do not follow the \cite{Cunha2010} SFR-{\mdust} relation, having higher dust masses for a given SFR. We also found that they evolve away from this relation as they age, with SFRs decreasing faster than dust masses.

We obtained a dust removal timescale for dusty elliptical galaxies of 2.26 $\pm$ 0.18 Gyr, which is consistent with the value of 2.5 $\pm$ 0.4 Gyr found by \cite{Michalowski2019}. The dust mass decline does not depend on stellar mass, implying a linear scaling of this effect with galaxy mass. Moreover there is no dependence of the decrease in dust mass on the galaxy environment or redshift, so the dust mass decline is of an internal nature. The independence of the dust decline on the central surface luminosity (a proxy for integrated black hole activity) suggests that AGN feedback is not responsible for the ISM decline.

\begin{acknowledgements}
A.L., M.J.M., J.N., and M.S.~acknowledge the support of 
the National Science Centre, Poland through the SONATA BIS grant 2018/30/E/ST9/00208.
This research was funded in whole or in part by National Science Centre, Poland (grant number: 2021/41/N/ST9/02662).
For the purpose of Open Access, the author has applied a CC-BY public copyright licence to any Author Accepted Manuscript (AAM) version arising from this submission.
A.L. and C.G.~acknowledge the support of the Leon Rosenfeld Foundation.
A.L.~acknowledges the support of Adam Mickiewicz University in Poznań, Poland via program Uniwersytet Jutra II (POWR.03.05.00-00-Z303/18).
O.R.~acknowledges the support of the National Science Centre, Poland through the grant 2022/01/4/ST9/00037.
This work is supported by a VILLUM FONDEN 
Investigator grant (project number 16599) and a Young Investigator Grant (project number 25501).

GAMA is a joint European-Australasian project based around a spectroscopic campaign using the Anglo-Australian Telescope. The GAMA input catalog is based on data taken from the Sloan Digital Sky Survey and the UKIRT Infrared Deep Sky Survey. Complementary imaging of the GAMA regions is being obtained by a number of independent survey programmes including GALEX MIS, VST KiDS, VISTA VIKING, WISE, Herschel-ATLAS, GMRT and ASKAP providing UV to radio coverage. GAMA is funded by the STFC (UK), the ARC (Australia), the AAO, and the participating institutions. The GAMA website is http://www.gama-survey.org/. 

\end{acknowledgements}

\bibliographystyle{aasjournal}
\bibliography{main}

\begin{thebibliography}{}
\expandafter\ifx\csname natexlab\endcsname\relax\def\natexlab#1{#1}\fi
\providecommand{\url}[1]{\href{#1}{#1}}

\bibitem[{{Agius} {et~al.}(2013){Agius}, {Sansom}, {Popescu}, {Andrae}, {Baes},
  {Baldry}, {Bourne}, {Brough}, {Clark}, {Conselice}, {Cooray}, {Dariush}, {De
  Zotti}, {Driver}, {Dunne}, {Eales}, {Foster}, {Gomez}, {H{\"a}u{\ss}ler},
  {Hopkins}, {Hopwood}, {Ivison}, {Kelvin}, {Lara-L{\'o}pez}, {Liske},
  {L{\'o}pez-S{\'a}nchez}, {Loveday}, {Maddox}, {Madore}, {Phillipps},
  {Robotham}, {Rowlands}, {Seibert}, {Smith}, {Temi}, {Tuffs}, \&
  {Valiante}}]{Agius2013}
{Agius}, N.~K., {Sansom}, A.~E., {Popescu}, C.~C., {et~al.} 2013, \mnras, 431,
  1929

\bibitem[{{Agius} {et~al.}(2015){Agius}, {di Serego Alighieri}, {Viaene},
  {Baes}, {Sansom}, {Bourne}, {Bland-Hawthorn}, {Brough}, {Davis}, {De Looze},
  {Driver}, {Dunne}, {Dye}, {Eales}, {Hughes}, {Ivison}, {Kelvin}, {Maddox},
  {Mahajan}, {Pappalardo}, {Robotham}, {Rowlands}, {Temi}, \&
  {Valiante}}]{Agius2015}
{Agius}, N.~K., {di Serego Alighieri}, S., {Viaene}, S., {et~al.} 2015, \mnras,
  451, 3815

\bibitem[{{Baldry} {et~al.}(2018){Baldry}, {Liske}, {Brown}, {Robotham},
  {Driver}, {Dunne}, {Alpaslan}, {Brough}, {Cluver}, {Eardley}, {Farrow},
  {Heymans}, {Hildebrandt}, {Hopkins}, {Kelvin}, {Loveday}, {Moffett},
  {Norberg}, {Owers}, {Taylor}, {Wright}, {Bamford}, {Bland -Hawthorn},
  {Bourne}, {Bremer}, {Colless}, {Conselice}, {Croom}, {Davies}, {Foster},
  {Grootes}, {Holwerda}, {Jones}, {Kafle}, {Kuijken}, {Lara-Lopez},
  {L{\'o}pez-S{\'a}nchez}, {Meyer}, {Phillipps}, {Sutherland}, {van Kampen}, \&
  {Wilkins}}]{Baldry2018}
{Baldry}, I.~K., {Liske}, J., {Brown}, M.~J.~I., {et~al.} 2018, MNRAS, 474,
  3875

\bibitem[{{Baldwin} {et~al.}(1981){Baldwin}, {Phillips}, \& {Terlevich}}]{bpt}
{Baldwin}, J.~A., {Phillips}, M.~M., \& {Terlevich}, R. 1981, \pasp, 93, 5

\bibitem[{{Balogh} {et~al.}(1999){Balogh}, {Morris}, {Yee}, {Carlberg}, \&
  {Ellingson}}]{Balogh1999}
{Balogh}, M.~L., {Morris}, S.~L., {Yee}, H.~K.~C., {Carlberg}, R.~G., \&
  {Ellingson}, E. 1999, \apj, 527, 54

\bibitem[{{Barlow}(1978)}]{Barlow1978}
{Barlow}, M.~J. 1978, MNRAS, 183, 367

\bibitem[{{Bianchi} \& {Ferrara}(2005)}]{Bianchi2005}
{Bianchi}, S., \& {Ferrara}, A. 2005, MNRAS, 358, 379

\bibitem[{{Bianchi} \& {Schneider}(2007)}]{Bianchi2007}
{Bianchi}, S., \& {Schneider}, R. 2007, MNRAS, 378, 973

\bibitem[{{Bl{\'a}nquez-Ses{\'e}} {et~al.}(2023){Bl{\'a}nquez-Ses{\'e}},
  {G{\'o}mez-Guijarro}, {Magdis}, {Magnelli}, {Gobat}, {Daddi}, {Franco},
  {Whitaker}, {Valentino}, {Adscheid}, {Schinnerer}, {Zanella}, {Xiao}, {Wang},
  {Liu}, {Kokorev}, \& {Elbaz}}]{Blanquez2023}
{Bl{\'a}nquez-Ses{\'e}}, D., {G{\'o}mez-Guijarro}, C., {Magdis}, G.~E.,
  {et~al.} 2023, arXiv e-prints, arXiv:2303.12110

\bibitem[{{Bluck} {et~al.}(2022){Bluck}, {Maiolino}, {Brownson}, {Conselice},
  {Ellison}, {Piotrowska}, \& {Thorp}}]{Bluck2022}
{Bluck}, A. F.~L., {Maiolino}, R., {Brownson}, S., {et~al.} 2022, \aap, 659,
  A160

\bibitem[{{Bluck} {et~al.}(2020{\natexlab{a}}){Bluck}, {Maiolino},
  {S{\'a}nchez}, {Ellison}, {Thorp}, {Piotrowska}, {Teimoorinia}, \&
  {Bundy}}]{Bluck2020}
{Bluck}, A. F.~L., {Maiolino}, R., {S{\'a}nchez}, S.~F., {et~al.}
  2020{\natexlab{a}}, \mnras, 492, 96

\bibitem[{{Bluck} {et~al.}(2023){Bluck}, {Piotrowska}, \&
  {Maiolino}}]{Bluck2023}
{Bluck}, A. F.~L., {Piotrowska}, J.~M., \& {Maiolino}, R. 2023, \apj, 944, 108

\bibitem[{{Bluck} {et~al.}(2020{\natexlab{b}}){Bluck}, {Maiolino},
  {Piotrowska}, {Trussler}, {Ellison}, {S{\'a}nchez}, {Thorp}, {Teimoorinia},
  {Moreno}, \& {Conselice}}]{Bluck2020b}
{Bluck}, A. F.~L., {Maiolino}, R., {Piotrowska}, J.~M., {et~al.}
  2020{\natexlab{b}}, \mnras, 499, 230

\bibitem[{{Bocchio} {et~al.}(2012){Bocchio}, {Micelotta}, {Gautier}, \&
  {Jones}}]{Bocchio2012}
{Bocchio}, M., {Micelotta}, E.~R., {Gautier}, A.~L., \& {Jones}, A.~P. 2012,
  \aap, 545, A124

\bibitem[{{Brough} {et~al.}(2013){Brough}, {Croom}, {Sharp}, {Hopkins},
  {Taylor}, {Baldry}, {Gunawardhana}, {Liske}, {Norberg}, {Robotham}, {Bauer},
  {Bland-Hawthorn}, {Colless}, {Foster}, {Kelvin}, {Lara-Lopez},
  {L{\'o}pez-S{\'a}nchez}, {Loveday}, {Owers}, {Pimbblet}, \&
  {Prescott}}]{Brough2013}
{Brough}, S., {Croom}, S., {Sharp}, R., {et~al.} 2013, \mnras, 435, 2903

\bibitem[{{Cardiel} {et~al.}(1998){Cardiel}, {Gorgas}, {Cenarro}, \&
  {Gonzalez}}]{Cardiel1998}
{Cardiel}, N., {Gorgas}, J., {Cenarro}, J., \& {Gonzalez}, J.~J. 1998, \aaps,
  127, 597

\bibitem[{{Chabrier}(2003)}]{Chabrier2003}
{Chabrier}, G. 2003, \apjl, 586, L133

\bibitem[{{Cherchneff} \& {Dwek}(2010)}]{Cherchneff2010}
{Cherchneff}, I., \& {Dwek}, E. 2010, ApJ, 713, 1

\bibitem[{{Conroy} {et~al.}(2015){Conroy}, {van Dokkum}, \&
  {Kravtsov}}]{Conroy2015}
{Conroy}, C., {van Dokkum}, P.~G., \& {Kravtsov}, A. 2015, ApJ, 803, 77

\bibitem[{{da Cunha} {et~al.}(2008){da Cunha}, {Charlot}, \&
  {Elbaz}}]{daCunha2008}
{da Cunha}, E., {Charlot}, S., \& {Elbaz}, D. 2008, \mnras, 388, 1595

\bibitem[{{da Cunha} {et~al.}(2010){da Cunha}, {Eminian}, {Charlot}, \&
  {Blaizot}}]{Cunha2010}
{da Cunha}, E., {Eminian}, C., {Charlot}, S., \& {Blaizot}, J. 2010, MNRAS,
  403, 1894

\bibitem[{{Dariush} {et~al.}(2016){Dariush}, {Dib}, {Hony}, {Smith},
  {Zhukovska}, {Dunne}, {Eales}, {Andrae}, {Baes}, {Baldry}, {Bauer},
  {Bland-Hawthorn}, {Brough}, {Bourne}, {Cava}, {Clements}, {Cluver}, {Cooray},
  {De Zotti}, {Driver}, {Grootes}, {Hopkins}, {Hopwood}, {Kaviraj}, {Kelvin},
  {Lara-Lopez}, {Liske}, {Loveday}, {Maddox}, {Madore}, {Micha{\l}owski},
  {Pearson}, {Popescu}, {Robotham}, {Rowlands}, {Seibert}, {Shabani}, {Smith},
  {Taylor}, {Tuffs}, {Valiante}, \& {Virdee}}]{Dariush2016}
{Dariush}, A., {Dib}, S., {Hony}, S., {et~al.} 2016, \mnras, 456, 2221

\bibitem[{{de Albernaz Ferreira} \& {Ferrari}(2018)}]{deAlbernazFerreira2018}
{de Albernaz Ferreira}, L., \& {Ferrari}, F. 2018, \mnras, 473, 2701

\bibitem[{{di Serego Alighieri} {et~al.}(2013){di Serego Alighieri}, {Bianchi},
  {Pappalardo}, {Zibetti}, {Auld}, {Baes}, {Bendo}, {Corbelli}, {Davies},
  {Davis}, {De Looze}, {Fritz}, {Gavazzi}, {Giovanardi}, {Grossi}, {Hunt},
  {Magrini}, {Pierini}, \& {Xilouris}}]{diseregoalighieri2013}
{di Serego Alighieri}, S., {Bianchi}, S., {Pappalardo}, C., {et~al.} 2013,
  \aap, 552, A8

\bibitem[{{Driver} {et~al.}(2011){Driver}, {Hill}, {Kelvin}, {Robotham},
  {Liske}, {Norberg}, {Baldry}, {Bamford}, {Hopkins}, {Loveday}, {Peacock},
  {Andrae}, {Bland-Hawthorn}, {Brough}, {Brown}, {Cameron}, {Ching}, {Colless},
  {Conselice}, {Croom}, {Cross}, {de Propris}, {Dye}, {Drinkwater}, {Ellis},
  {Graham}, {Grootes}, {Gunawardhana}, {Jones}, {van Kampen}, {Maraston},
  {Nichol}, {Parkinson}, {Phillipps}, {Pimbblet}, {Popescu}, {Prescott},
  {Roseboom}, {Sadler}, {Sansom}, {Sharp}, {Smith}, {Taylor}, {Thomas},
  {Tuffs}, {Wijesinghe}, {Dunne}, {Frenk}, {Jarvis}, {Madore}, {Meyer},
  {Seibert}, {Staveley-Smith}, {Sutherland}, \& {Warren}}]{Driver2011}
{Driver}, S.~P., {Hill}, D.~T., {Kelvin}, L.~S., {et~al.} 2011, \mnras, 413,
  971

\bibitem[{{Driver} {et~al.}(2016){Driver}, {Wright}, {Andrews}, {Davies},
  {Kafle}, {Lange}, {Moffett}, {Mannering}, {Robotham}, {Vinsen}, {Alpaslan},
  {Andrae}, {Baldry}, {Bauer}, {Bamford}, {Bland-Hawthorn}, {Bourne}, {Brough},
  {Brown}, {Cluver}, {Croom}, {Colless}, {Conselice}, {da Cunha}, {De Propris},
  {Drinkwater}, {Dunne}, {Eales}, {Edge}, {Frenk}, {Graham}, {Grootes},
  {Holwerda}, {Hopkins}, {Ibar}, {van Kampen}, {Kelvin}, {Jarrett}, {Jones},
  {Lara-Lopez}, {Liske}, {Lopez-Sanchez}, {Loveday}, {Maddox}, {Madore},
  {Mahajan}, {Meyer}, {Norberg}, {Penny}, {Phillipps}, {Popescu}, {Tuffs},
  {Peacock}, {Pimbblet}, {Prescott}, {Rowlands}, {Sansom}, {Seibert}, {Smith},
  {Sutherland}, {Taylor}, {Valiante}, {Vazquez-Mata}, {Wang}, {Wilkins}, \&
  {Williams}}]{Driver2016}
{Driver}, S.~P., {Wright}, A.~H., {Andrews}, S.~K., {et~al.} 2016, MNRAS, 455,
  3911

\bibitem[{{Dwek} \& {Scalo}(1980)}]{Dwek1980}
{Dwek}, E., \& {Scalo}, J.~M. 1980, ApJ, 239, 193

\bibitem[{{Fabian}(2012)}]{Fabian2012}
{Fabian}, A.~C. 2012, \araa, 50, 455

\bibitem[{{Gall} {et~al.}(2011){Gall}, {Andersen}, \& {Hjorth}}]{Gall2011}
{Gall}, C., {Andersen}, A.~C., \& {Hjorth}, J. 2011, A$\&$A, 528, A14

\bibitem[{{Gall} \& {Hjorth}(2018)}]{Gall2018}
{Gall}, C., \& {Hjorth}, J. 2018, ApJ, 868, 62

\bibitem[{{Griffin} {et~al.}(2010){Griffin}, {Abergel}, {Abreu}, {Ade},
  {Andr{\'e}}, {Augueres}, {Babbedge}, {Bae}, {Baillie}, {Baluteau}, {Barlow},
  {Bendo}, {Benielli}, {Bock}, {Bonhomme}, {Brisbin}, {Brockley-Blatt},
  {Caldwell}, {Cara}, {Castro-Rodriguez}, {Cerulli}, {Chanial}, {Chen},
  {Clark}, {Clements}, {Clerc}, {Coker}, {Communal}, {Conversi}, {Cox},
  {Crumb}, {Cunningham}, {Daly}, {Davis}, {de Antoni}, {Delderfield}, {Devin},
  {di Giorgio}, {Didschuns}, {Dohlen}, {Donati}, {Dowell}, {Dowell}, {Duband},
  {Dumaye}, {Emery}, {Ferlet}, {Ferrand}, {Fontignie}, {Fox}, {Franceschini},
  {Frerking}, {Fulton}, {Garcia}, {Gastaud}, {Gear}, {Glenn}, {Goizel},
  {Griffin}, {Grundy}, {Guest}, {Guillemet}, {Hargrave}, {Harwit}, {Hastings},
  {Hatziminaoglou}, {Herman}, {Hinde}, {Hristov}, {Huang}, {Imhof}, {Isaak},
  {Israelsson}, {Ivison}, {Jennings}, {Kiernan}, {King}, {Lange}, {Latter},
  {Laurent}, {Laurent}, {Leeks}, {Lellouch}, {Levenson}, {Li}, {Li},
  {Lilienthal}, {Lim}, {Liu}, {Lu}, {Madden}, {Mainetti}, {Marliani}, {McKay},
  {Mercier}, {Molinari}, {Morris}, {Moseley}, {Mulder}, {Mur}, {Naylor},
  {Nguyen}, {O'Halloran}, {Oliver}, {Olofsson}, {Olofsson}, {Orfei}, {Page},
  {Pain}, {Panuzzo}, {Papageorgiou}, {Parks}, {Parr-Burman}, {Pearce},
  {Pearson}, {P{\'e}rez-Fournon}, {Pinsard}, {Pisano}, {Podosek}, {Pohlen},
  {Polehampton}, {Pouliquen}, {Rigopoulou}, {Rizzo}, {Roseboom}, {Roussel},
  {Rowan-Robinson}, {Rownd}, {Saraceno}, {Sauvage}, {Savage}, {Savini},
  {Sawyer}, {Scharmberg}, {Schmitt}, {Schneider}, {Schulz}, {Schwartz},
  {Shafer}, {Shupe}, {Sibthorpe}, {Sidher}, {Smith}, {Smith}, {Smith},
  {Spencer}, {Stobie}, {Sudiwala}, {Sukhatme}, {Surace}, {Stevens}, {Swinyard},
  {Trichas}, {Tourette}, {Triou}, {Tseng}, {Tucker}, {Turner}, {Vaccari},
  {Valtchanov}, {Vigroux}, {Virique}, {Voellmer}, {Walker}, {Ward}, {Waskett},
  {Weilert}, {Wesson}, {White}, {Whitehouse}, {Wilson}, {Winter}, {Woodcraft},
  {Wright}, {Xu}, {Zavagno}, {Zemcov}, {Zhang}, \& {Zonca}}]{spire}
{Griffin}, M.~J., {Abergel}, A., {Abreu}, A., {et~al.} 2010, \aap, 518, L3

\bibitem[{{H{\"a}ring} \& {Rix}(2004)}]{Haring2004}
{H{\"a}ring}, N., \& {Rix}, H.-W. 2004, \apjl, 604, L89

\bibitem[{{Hjorth} {et~al.}(2014){Hjorth}, {Gall}, \&
  {Micha{\l}owski}}]{Hjorth2014}
{Hjorth}, J., {Gall}, C., \& {Micha{\l}owski}, M.~J. 2014, \apjl, 782, L23

\bibitem[{{Kelvin} {et~al.}(2012){Kelvin}, {Driver}, {Robotham}, {Hill},
  {Alpaslan}, {Baldry}, {Bamford}, {Bland-Hawthorn}, {Brough}, {Graham},
  {H{\"a}ussler}, {Hopkins}, {Liske}, {Loveday}, {Norberg}, {Phillipps},
  {Popescu}, {Prescott}, {Taylor}, \& {Tuffs}}]{Kelvin2012}
{Kelvin}, L.~S., {Driver}, S.~P., {Robotham}, A. S.~G., {et~al.} 2012, \mnras,
  421, 1007

\bibitem[{{Laki{\'c}evi{\'c}} {et~al.}(2015){Laki{\'c}evi{\'c}}, {van Loon},
  {Meixner}, {Gordon}, {Bot}, {Roman-Duval}, {Babler}, {Bolatto},
  {Engelbracht}, {Filipovi{\'c}}, {Hony}, {Indebetouw}, {Misselt}, {Montiel},
  {Okumura}, {Panuzzo}, {Patat}, {Sauvage}, {Seale}, {Sonneborn}, {Temim},
  {Uro{\v s}evi{\'c}}, \& {Zanardo}}]{Lakicevic2015}
{Laki{\'c}evi{\'c}}, M., {van Loon}, J.~T., {Meixner}, M., {et~al.} 2015, ApJ,
  799, 50

\bibitem[{{Li} {et~al.}(2020){Li}, {Li}, {Bryan}, {Ostriker}, \&
  {Quataert}}]{Li2020}
{Li}, M., {Li}, Y., {Bryan}, G.~L., {Ostriker}, E.~C., \& {Quataert}, E. 2020,
  \apj, 898, 23

\bibitem[{{Magdis} {et~al.}(2021){Magdis}, {Gobat}, {Valentino}, {Daddi},
  {Zanella}, {Kokorev}, {Toft}, {Jin}, \& {Whitaker}}]{Magdis2021}
{Magdis}, G.~E., {Gobat}, R., {Valentino}, F., {et~al.} 2021, \aap, 647, A33

\bibitem[{{Magorrian} {et~al.}(1998){Magorrian}, {Tremaine}, {Richstone},
  {Bender}, {Bower}, {Dressler}, {Faber}, {Gebhardt}, {Green}, {Grillmair},
  {Kormendy}, \& {Lauer}}]{Magorrian1998}
{Magorrian}, J., {Tremaine}, S., {Richstone}, D., {et~al.} 1998, \aj, 115, 2285

\bibitem[{{Micha{\l}owski} {et~al.}(2019){Micha{\l}owski}, {Hjorth}, {Gall},
  {Frayer}, {Tsai}, {Hirashita}, {Rowland s}, {Takeuchi}, {Le{\'s}niewska},
  {Behrendt}, {Bourne}, {Hughes}, {Spring}, {Zavala}, \&
  {Bartczak}}]{Michalowski2019}
{Micha{\l}owski}, M.~J., {Hjorth}, J., {Gall}, C., {et~al.} 2019, A$\&$A, 632,
  A43

\bibitem[{{Nadolny} {et~al.}(in prep.){Nadolny}, {Micha{\l}owski}, \&
  {Le{\'s}niewska}}]{Nadolny2023}
{Nadolny}, J., {Micha{\l}owski}, M.~J., \& {Le{\'s}niewska}, A. in prep.,
  \apjl, in prep.

\bibitem[{{Pilbratt} {et~al.}(2010){Pilbratt}, {Riedinger}, {Passvogel},
  {Crone}, {Doyle}, {Gageur}, {Heras}, {Jewell}, {Metcalfe}, {Ott}, \&
  {Schmidt}}]{Pilbratt2010}
{Pilbratt}, G.~L., {Riedinger}, J.~R., {Passvogel}, T., {et~al.} 2010, \aap,
  518, L1

\bibitem[{{Piotrowska} {et~al.}(2022){Piotrowska}, {Bluck}, {Maiolino}, \&
  {Peng}}]{Piotrowska2022}
{Piotrowska}, J.~M., {Bluck}, A. F.~L., {Maiolino}, R., \& {Peng}, Y. 2022,
  \mnras, 512, 1052

\bibitem[{{Rowlands} {et~al.}(2012){Rowlands}, {Dunne}, {Maddox}, {Bourne},
  {Gomez}, {Kaviraj}, {Bamford}, {Brough}, {Charlot}, {da Cunha}, {Driver},
  {Eales}, {Hopkins}, {Kelvin}, {Nichol}, {Sansom}, {Sharp}, {Smith}, {Temi},
  {van der Werf}, {Baes}, {Cava}, {Cooray}, {Croom}, {Dariush}, {de Zotti},
  {Dye}, {Fritz}, {Hopwood}, {Ibar}, {Ivison}, {Liske}, {Loveday}, {Madore},
  {Norberg}, {Popescu}, {Rigby}, {Robotham}, {Rodighiero}, {Seibert}, \&
  {Tuffs}}]{Rowlands2012}
{Rowlands}, K., {Dunne}, L., {Maddox}, S., {et~al.} 2012, MNRAS, 419, 2545

\bibitem[{{Ryzhov} {et~al.}(in prep.){Ryzhov}, {Micha{\l}owski},
  {Le{\'s}niewska}, \& {Nadolny}}]{Ryzhov2023}
{Ryzhov}, O., {Micha{\l}owski}, M.~J., {Le{\'s}niewska}, A., \& {Nadolny}, J.
  in prep., \apjl, in prep.

\bibitem[{{Scoville} {et~al.}(2016){Scoville}, {Sheth}, {Aussel}, {Vanden
  Bout}, {Capak}, {Bongiorno}, {Casey}, {Murchikova}, {Koda},
  {{\'A}lvarez-M{\'a}rquez}, {Lee}, {Laigle}, {McCracken}, {Ilbert}, {Pope},
  {Sanders}, {Chu}, {Toft}, {Ivison}, \& {Manohar}}]{Scoville2016}
{Scoville}, N., {Sheth}, K., {Aussel}, H., {et~al.} 2016, \apj, 820, 83

\bibitem[{{S{\'e}rsic}(1963)}]{Sersic1963}
{S{\'e}rsic}, J.~L. 1963, Boletin de la Asociacion Argentina de Astronomia La
  Plata Argentina, 6, 41

\bibitem[{{Slavin} {et~al.}(2015){Slavin}, {Dwek}, \& {Jones}}]{Slavin2015}
{Slavin}, J.~D., {Dwek}, E., \& {Jones}, A.~P. 2015, ApJ, 803, 7

\bibitem[{{Smith} {et~al.}(2011){Smith}, {Dunne}, {Maddox}, {Eales},
  {Bonfield}, {Jarvis}, {Sutherland}, {Fleuren}, {Rigby}, {Thompson}, {Baldry},
  {Bamford}, {Buttiglione}, {Cava}, {Clements}, {Cooray}, {Croom}, {Dariush},
  {de Zotti}, {Driver}, {Dunlop}, {Fritz}, {Hill}, {Hopkins}, {Hopwood},
  {Ibar}, {Ivison}, {Jones}, {Kelvin}, {Leeuw}, {Liske}, {Loveday}, {Madore},
  {Norberg}, {Panuzzo}, {Pascale}, {Pohlen}, {Popescu}, {Prescott}, {Robotham},
  {Rodighiero}, {Scott}, {Seibert}, {Sharp}, {Temi}, {Tuffs}, {van der Werf},
  \& {van Kampen}}]{Smith2011}
{Smith}, D.~J.~B., {Dunne}, L., {Maddox}, S.~J., {et~al.} 2011, \mnras, 416,
  857

\bibitem[{{Smith} {et~al.}(2012){Smith}, {Gomez}, {Eales}, {Ciesla}, {Boselli},
  {Cortese}, {Bendo}, {Baes}, {Bianchi}, {Clemens}, {Clements}, {Cooray},
  {Davies}, {De Looze}, {di Serego Alighieri}, {Fritz}, {Gavazzi}, {Gear},
  {Madden}, {Mentuch}, {Panuzzo}, {Pohlen}, {Spinoglio}, {Verstappen},
  {Vlahakis}, {Wilson}, \& {Xilouris}}]{Smith2012}
{Smith}, M.~W.~L., {Gomez}, H.~L., {Eales}, S.~A., {et~al.} 2012, \apj, 748,
  123

\bibitem[{{Speagle} {et~al.}(2014){Speagle}, {Steinhardt}, {Capak}, \&
  {Silverman}}]{Speagle2014}
{Speagle}, J.~S., {Steinhardt}, C.~L., {Capak}, P.~L., \& {Silverman}, J.~D.
  2014, ApJS, 214, 15

\bibitem[{{Temim} {et~al.}(2015){Temim}, {Dwek}, {Tchernyshyov}, {Boyer},
  {Meixner}, {Gall}, \& {Roman-Duval}}]{Temim2015}
{Temim}, T., {Dwek}, E., {Tchernyshyov}, K., {et~al.} 2015, ApJ, 799, 158

\bibitem[{{Whitaker} {et~al.}(2021){Whitaker}, {Williams}, {Mowla}, {Spilker},
  {Toft}, {Narayanan}, {Pope}, {Magdis}, {van Dokkum}, {Akhshik}, {Bezanson},
  {Brammer}, {Leja}, {Man}, {Nelson}, {Richard}, {Pacifici}, {Sharon}, \&
  {Valentino}}]{Whitaker2021}
{Whitaker}, K.~E., {Williams}, C.~C., {Mowla}, L., {et~al.} 2021, \nat, 597,
  485

\end{thebibliography}

\newpage
\appendix
\section{Numerical values from figures}

\begin{table*}[h!]
\caption{Dust removal timescale $\tau$, half life-times, and normalisations from fitting exponential functions to the middle panel of Figure \ref{Fig1} and all panels from Figure \ref{Fig2}.}
\begin{center}
\begin{tabular}{c|ccc}
 & $\tau$ [Gyr] & $\tau_{1/2}$ [Gyr] & $A$ \\
 \hline\hline
 MS galaxies & 1.53 $\pm$ 0.22 & 1.06 $\pm$ 0.15 & $-$2.16 $\pm$ 0.04\\
 below-MS & 2.36 $\pm$ 0.22 & 1.64 $\pm$ 0.15 & $-$2.33 $\pm$ 0.03\\
 all galaxies & 2.26 $\pm$ 0.18 & 1.57 $\pm$ 0.12 & $-$2.31 $\pm$ 0.02\\
\hline\hline
$log(M_{stellar}/M_{\odot})$ \\
10.0 -- 10.5 & 2.71 $\pm$ 0.72 & 1.88 $\pm$ 0.50 & $-$2.34 $\pm$ 0.05\\
10.5 -- 11.0 & 3.03 $\pm$ 0.40 & 2.10 $\pm$ 0.27 & $-$2.51 $\pm$ 0.03\\
11.0 -- 11.5 & 2.98 $\pm$ 0.59 & 2.07 $\pm$ 0.41 & $-$2.56 $\pm$ 0.05\\
\hline
$\Sigma / M pc^{-2}$ \\
$<$ 0.3 & 2.40 $\pm$ 0.81 & 1.66 $\pm$ 0.56 & $-$2.45 $\pm$ 0.08\\
0.3 -- 0.8 & 2.36 $\pm$ 0.87 & 1.63 $\pm$ 0.60 & $-$2.48 $\pm$ 0.10\\
$>$ 0.8 & 1.97 $\pm$ 0.57 & 1.36 $\pm$ 0.40 & $-$2.33 $\pm$ 0.11\\
\hline
$z$ \\
0.01 -- 0.12 & 2.02 $\pm$ 0.28 & 1.40 $\pm$ 0.19 & $-$2.38 $\pm$ 0.05\\
0.12 -- 0.25 & 2.64 $\pm$ 0.33 & 1.83 $\pm$ 0.23 & $-$2.37 $\pm$ 0.03\\
0.25 -- 0.32 & 2.64 $\pm$ 0.52 & 1.83 $\pm$ 0.36 & $-$2.31 $\pm$ 0.04\\
\hline\hline
\end{tabular}
\end{center}
\label{tab1}
\end{table*}

\begin{table*}[h!]
\caption{Medians of SFRs, stellar masses, dust masses (standard deviation in brackets) plotted in figure \ref{Fig3}, and number of galaxies for each age bin.}
\begin{tabular}{c|cccc|cccc}
& & MS galaxies & & & & below-MS & & \\
\hline\hline
log(age) & log(SFR) & log(M$_{stellar}$) & log(M$_{dust}$) & number & log(SFR) & log(M$_{stellar}$) & log(M$_{dust}$) & number \\
$[$yr] & [M$_\odot$ yr$^{-1}$] & [M$_\odot$] & [M$_\odot$] & of galaxies & [M$_\odot$ yr$^{-1}$] & [M$_\odot$] & [M$_\odot$] & of galaxies\\
\hline\hline
$<$ 9.0 & 0.80 (0.56) & 10.15 (0.61) & 7.76 (0.54) & 133 & $-$0.64 (0.91) & 9.95 (1.02) & 7.63 (0.84) & 15\\
9.0 -- 9.2 & 0.72 (0.44) & 10.48 (0.45) & 7.78 (0.50) & 151 & $-$0.04 (0.43) & 10.57 (0.51) & 7.88 (0.56) & 37\\
9.2 -- 9.4 & 0.69 (0.39) & 10.65 (0.39) & 7.87 (0.40) & 266 & $-$0.04 (0.48) & 10.71 (0.46) & 7.87 (0.48) & 171\\
9.4 -- 9.5 & 0.53 (0.30) & 10.72 (0.35) & 7.77 (0.43) & 119 & 0.04 (0.47) & 10.78 (0.34) & 7.93 (0.50) & 207\\
9.5 -- 9.6 & 0.47 (0.22) & 10.81 (0.20) & 7.67 (0.37) & 33 & $-$0.07 (0.48) & 10.84 (0.33) & 7.81 (0.47) & 301\\
9.6 -- 9.7 & 0.63 (0.47) & 10.92 (0.31) & 7.57 (0.37) & 17 & $-$0.27 (0.48) & 10.93 (0.29) & 7.77 (0.45) & 328\\
9.7 -- 9.8 & 0.52 (0.21) & 10.66 (0.11) & 7.94 (0.58) & 3 & $-$0.58 (0.54) & 10.98 (0.31) & 7.56 (0.52) & 198\\
9.8 -- 10.0 & $\cdots$ & $\cdots$ & $\cdots$ & 0 & $-$0.97 (0.49) & 10.96 (0.28) & 7.32 (0.57) & 71\\
\hline\hline
\end{tabular}
\label{tab2}
\tablebib{Median and standard deviation calculated based on logarithmic values of SFR, M$_{stellar}$, and M$_{dust}$ parameters in each bin.}
\end{table*}

\end{document}